\documentclass[10pt,a4paper]{iopart}

\usepackage[T1]{fontenc}
\usepackage[latin9]{inputenc}

\expandafter\let\csname equation*\endcsname\relax
\expandafter\let\csname endequation*\endcsname\relax
\usepackage{amsmath}

\usepackage{graphicx}
\usepackage{booktabs}

\begin{document}

\title{Environment-assisted quantum transport in ordered systems}

\author{Ivan Kassal$^{1,2}$ and Al\'an Aspuru-Guzik$^2$}

\address{$^1$ School of Mathematics and Physics and Centre for Engineered Quantum
Systems,\\ The University of Queensland, St Lucia QLD 4072, Australia}

\address{$^2$ Department of Chemistry and Chemical Biology,\\ Harvard University,
Cambridge MA 02138, United States}

\ead{i.kassal@uq.edu.au}

\begin{abstract}
Noise-assisted transport in quantum systems occurs when quantum time-evolution
and decoherence conspire to produce a transport efficiency that is
higher than what would be seen in either the purely quantum or purely
classical cases. In disordered systems, it has been understood as
the suppression of coherent quantum localisation through noise, which
brings detuned quantum levels into resonance and thus facilitates
transport. We report several new mechanisms of environment-assisted
transport in ordered systems, in which there is no localisation to overcome 
and where one would naively expect that coherent transport is the fastest possible. 
Although we are particularly motivated by the need to understand excitonic energy transfer
in photosynthetic light-harvesting complexes, our model is general---transport
in a tight-binding system with dephasing, a source, and a trap---and can be expected
to have wider application.
\end{abstract}


\maketitle

Recent experimental studies of photosynthetic light-harvesting complexes
have confronted us with the fact that at least some of these systems
exhibit excitonic coherence that is surprisingly long considering
their noisy environment \cite{Engel:2007hb,Panitchayangkoon:2010fw,Collini:2010hb}.
This makes it clear that if we are to understand their high light-harvesting
efficiency, we must study the ways in which quantum transport is affected
by the interplay of coherence and noise \cite{Rebentrost:2009hu,Plenio:2008ff,Gaab:2004hq,Rebentrost:2009tv,Caruso:2009ib,OlayaCastro:2008jz,Mohseni:2008gp,Lloyd:2011wg,Cao:2009vs,Perdomo:2010di,Vlaming:2007jk,Mohseni:2011wg,CamposVenuti:2011di,Hoyer:2010fl,n1,n2,n3,n4}.
It has been found that noise can enhance quantum transport in model
excitonic Hamiltonians \cite{Gaab:2004hq,Rebentrost:2009hu,Plenio:2008ff},
a phenomenon called environment-assisted quantum transport (ENAQT)
or decoherence-assisted transport. 

In the simplest approach, different environments around each chromophore
lead to a tight-binding model with sites that have different energies
(disorder). Because of disorder, the exciton becomes localised through
coherent phenomena such as destructive interference or Anderson localisation
\cite{Rebentrost:2009hu,Plenio:2008ff,Caruso:2009ib,Lloyd:2011wg}.
ENAQT is then simple to understand: noise can destroy the coherent
localisation, helping the exciton reach the trap site and increasing
the efficiency. Alternatively, decoherence has been described as fluctuations
of site energies which can transiently bring levels into resonance,
facilitating transport. 

If these interpretations were the whole story, ENAQT would be impossible
in ordered systems, those without energetic disorder. The absence
of ENAQT in ordered linear chains was predicted at least twice: for
example, Cao and Silbey predict ``the lack of environment-assistance in
linear-chain systems'' \cite{Cao:2009vs} while Plenio and Huelga note
``the expectation that noise does not enhance the transport of
excitations'' in uniform chains \cite{Plenio:2008ff}.
This expectation is strengthened by proofs of the impossibility of ENAQT in end-to-end transport
in ordered chains \cite{Plenio:2008ff,Cao:2009vs}. We revisit the
ordered chain and show that the case of end-to-end transport is the
\emph{only} case where ENAQT is impossible: its absence in end-to-end
transport is the rare exception that we are able to explain. We anticipate
that these findings will shed light on the efficiency of transport
in ordered excitonic systems, whether artificial, such as \emph{J}-aggregates,
or natural, such as the LHII complex in purple bacteria and the chlorosome
in green sulphur bacteria.

Results related to ours were reported by Gaab and Bardeen \cite{Gaab:2004hq}.
They considered ordered systems and noted that the environment can
sometimes enhance the {}``effective trapping rate.'' By contrast,
we focus on the trapping efficiency, a measure of how often the exciton
is productively trapped as opposed to lost, regardless of how fast
the transport is (loss is not modelled in \cite{Gaab:2004hq}, so
the efficiency is always 1). We do not average over initial sites,
which allows us to explain the absence of ENAQT in end-to-end transport.

\section{The model and the definition of ENAQT}

The ordered system we consider is a one-dimensional array of $N$
identical sites, coupled to their nearest neighbours and described
by the Hamiltonian
\begin{equation}
H_{0}=V\sum_{m=1}^{N-1}\left(|m\rangle\langle m+1|+|m+1\rangle\langle m|\right),\label{eq:H}
\end{equation}
where $V$ is the coupling strength and $\hbar=1$. This Hamiltonian
is equivalent to a system of coupled two-level systems that is restricted
to the single-excitation sector.

To study the efficiency of transport mediated by this Hamiltonian,
we introduce two distinct attenuation mechanisms. First, the particle
is irreversibly \emph{lost} from each site at an equal rate $\mu$,
modelling processes such as exciton recombination. Second, at a particular
trap site $\left|\tau\right\rangle $, the particle can be \emph{trapped}
at a rate $\kappa$, modelling, for example, the transfer of an exciton
to a photosynthetic reaction centre. These attenuation mechanisms
are incorporated by adding a non-hermitian part to the Hamiltonian,
\begin{equation}
H_{\mathrm{atten}}=-i\mu\sum_{m}|m\rangle\langle m|-i\kappa\left|\tau\right\rangle \left\langle \tau\right|.
\end{equation}
Loss and trapping both result in particle disappearance and have the
same mathematical form; the distinction is that we consider the energy
carried by lost particles to be unavailable while the trapped energy
to be productively useable. The norm of the state at time $t$ is
the probability that the particle will survive that long.

\begin{figure}
\textbf{a)}\includegraphics[width=8.5cm]{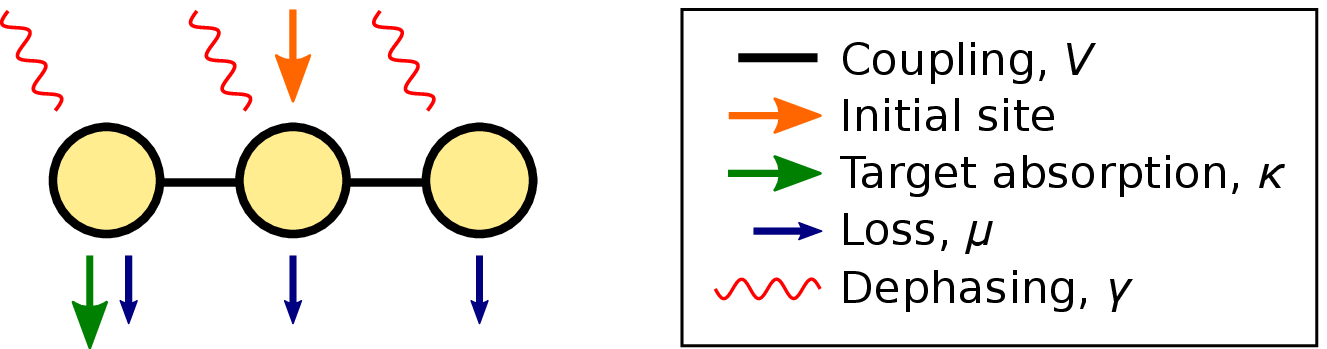}

\bigskip{}

\textbf{b)}\includegraphics[width=8.5cm]{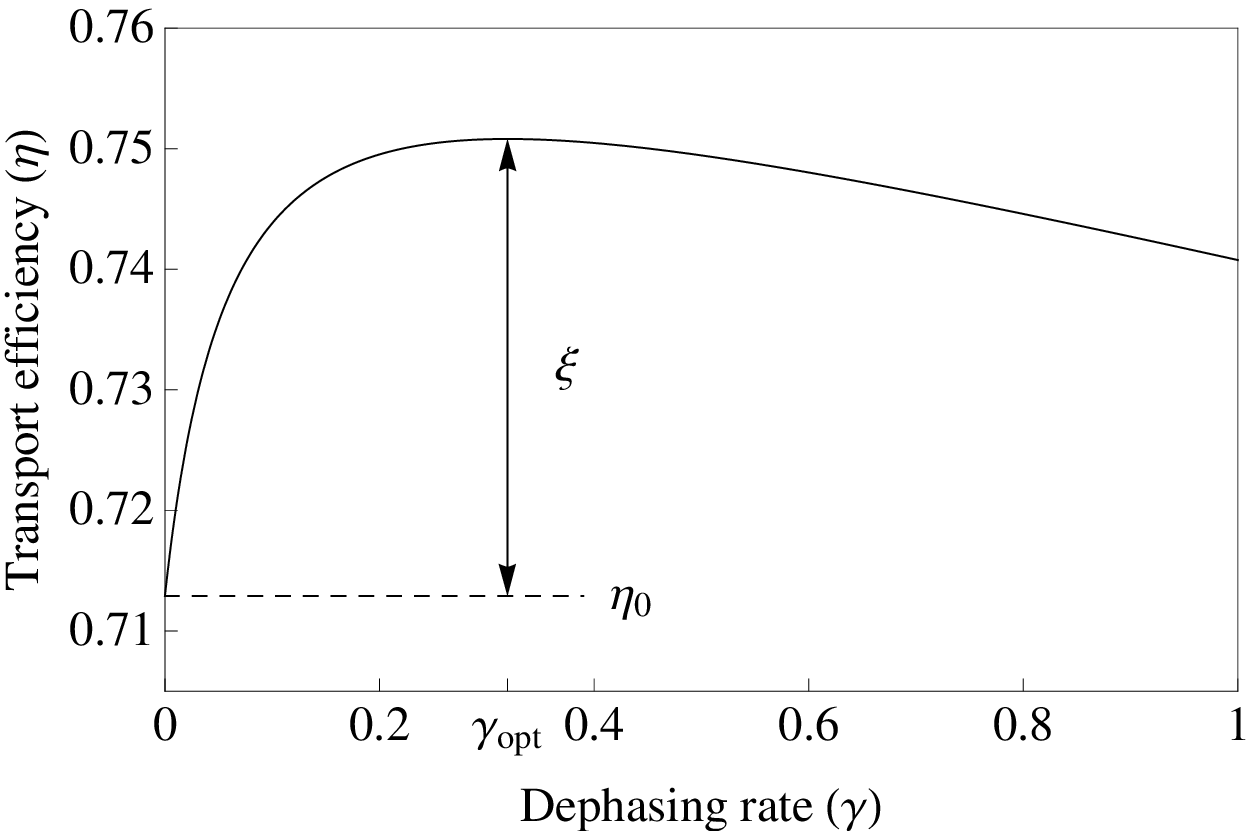}

\caption{Definition of ENAQT. \textbf{a)} Three-site chain with initial site
$\left|2\right\rangle $ and trap site $\left|1\right\rangle $. \textbf{b)
}Transport efficiency $\eta$ in the three-site chain, with trapping
rate $\kappa=0.1$ and loss rate $\mu=0.01$. $\eta$ is maximised
at the optimal dephasing rate $\gamma_{\mathrm{opt}}=0.319$. ENAQT
is the magnitude of the enhancement, $\xi=\eta_{\mathrm{max}}-\eta_{0}$,
or, in this case, $\xi=0.038$. \label{fig:Definition-of-ENAQT}}
$ $
\end{figure}

The attenuation mechanisms continuously reduces the particle's survival
probability, so that after a sufficiently long time, $t\gg\mu^{-1}$,
the probability of finding the particle is negligible. If $\rho(t)$
is the system's density matrix at time $t$, the probability of trapping
the particle in the interval $\left[t,t+dt\right]$ is $2\kappa\left\langle \tau\left|\rho(t)\right|\tau\right\rangle dt$.
The efficiency of transport, for initial state $\rho(0)$, is then
the overall trapping probability,
\begin{equation}
\eta=2\kappa\int_{0}^{\infty}\langle\tau|\rho(t)|\tau\rangle\, dt.
\end{equation}
Likewise, the probability of loss is $\eta'=2\mu\sum_{m}\int_{0}^{\infty}\langle m|\rho(t)|m\rangle\, dt=2\mu\int_{0}^{\infty}\mathrm{tr}\,\rho(t)\, dt$
and these branching ratios satisfy $\eta+\eta'=1$.

Environmental effects are modelled as (Markovian) pure dephasing,
acting independently on all sites with an equal rate $\gamma$. We
choose dephasing because it is one of the simplest forms of noise,
giving us a single-parameter minimal model that exhibits the desired
behaviour. Insofar as dephasing is the appropriate limit of several
more realistic noise models, we can expect qualitatively similar effects
if more complicated environments are considered. The dephasing superoperator
$\mathcal{D}$ is defined through $\left(\mathcal{D}\rho\right)_{nm}=-2\gamma\left(1-\delta_{nm}\right)\rho_{nm}$
and the resulting complete equation of motion is
\begin{equation}
\dot{\rho}=\mathcal{L}\rho=-i\left(H\rho-\rho H^{\dagger}\right)+\mathcal{D}\rho,\label{eq:master}
\end{equation}
where the total Hamiltonian is $H=H_{0}+H_{\mathrm{atten}}$. We note
that this master equation has been solved exactly for the case $\kappa=\mu=0$
both in the single-particle situation \cite{Esposito:2005dz,Esposito:2005fv}
and in the non-equilibrium setting for many particles \cite{Znidaric:2010wk,Znidaric:2011dp}.
In the following, we give a method for the exact (single-particle)
solution for any chain length and any $\kappa$ and $\mu$, allowing
us to calculate the efficiency. 

The coupling $V$ sets the energy scale, so we can take $V=1$. Then,
at every choice of loss and trapping rates $\mu$ and $\kappa$, the
efficiency $\eta$ is a function of the dephasing rate $\gamma$,
see Fig. \ref{fig:Definition-of-ENAQT}b. We observe that if $\gamma$
is very large, the particle will be localised at its initial site
due to the Zeno effect. Therefore, it will not be able to reach the
trap before it is lost, meaning that $\eta\to0$ as $\gamma\to\infty$.
Consequently, the maximum transport efficiency $\eta_{\mathrm{max}}$
will occur at a finite $\gamma_{\mathrm{opt}}\ge0$. ENAQT occurs
if $\gamma_{\mathrm{opt}}>0$, and is defined to be
\begin{equation}
\xi(\mu,\kappa)=\eta_{\mathrm{max}}-\eta_{0}.
\end{equation}

\begin{figure*}
{\footnotesize Efficiency without dephasing ($\eta_{0}$)\hfill ENAQT ($\xi$)\hfill Optimal
dephasing ($\gamma_{\mathrm{opt}}$)}

\includegraphics[width=1\textwidth]{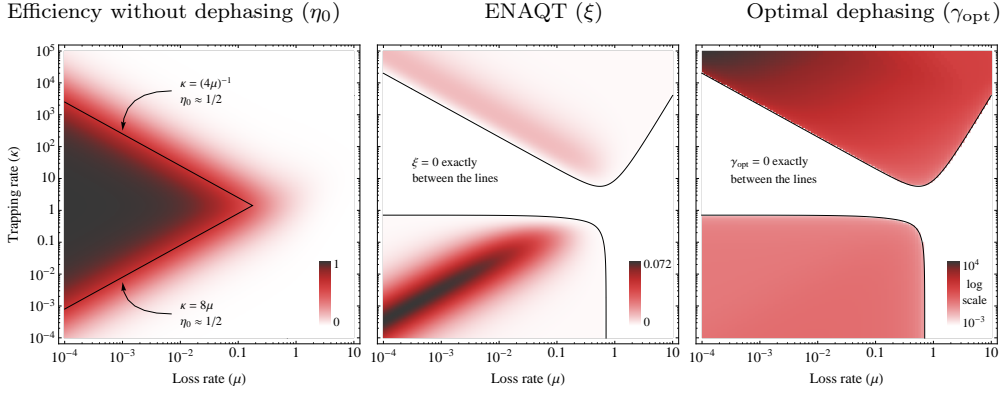}

\caption{Characterisation of ENAQT in the three-site system from Fig. \ref{fig:Definition-of-ENAQT}a.
Each of the three parameters defined in Fig. \ref{fig:Definition-of-ENAQT}b
($\eta_{0}$, $\xi$, and $\gamma_{\mathrm{opt}}$) is displayed as
a function of trapping and loss rates. \textbf{a)} High efficiency
in the regime with no dephasing is possible only for small loss and
intermediate trapping. The two lines indicate the region where the
efficiency is around $1/2$, meaning that neither loss nor trapping
dominates. They are computed from Eq. \ref{eq:etadef} ($\eta_{0}=\alpha_{0}/\beta_{0}$).
\textbf{b)} Large ENAQT occurs in regions not far from the lines in
\textbf{a}, where dephasing can push the balance between loss and
trapping in favour of trapping. \textbf{c)} In the lower left, $\xi_{\mathrm{max}}$
occurs when $\gamma_{\mathrm{opt}}\approx0.4151$. In the upper left,
strong trapping presents a high potential barrier, meaning that strong
dephasing is necessary for ENAQT. \label{fig:ENAQT-finite}}
\end{figure*}

We can now consider efficiency and ENAQT as a function of $\kappa$
and $\mu$. The following descriptions are all borne out in the example
in Fig. \ref{fig:ENAQT-finite}, which shows $\eta_{0}$, $\xi$,
and $\gamma_{\mathrm{opt}}$ as a function of $\kappa$ and $\mu$
in the finite system of three sites with the trap at one end and the
initial site in the middle (see Fig. \ref{fig:Definition-of-ENAQT}a).
Regardless of the number of sites, several limiting cases can be easily
understood. First, if $\kappa\ll\mu$, the particle will be lost before
it can be trapped, regardless of the amount of dephasing present.
Second, at large $\kappa$, the Hamiltonian term $-i\kappa\left|\tau\right\rangle \left\langle \tau\right|$
presents a high potential barrier for the particle, meaning that it
will be largely unable to access the trapping site (this is the case
even though the potential is imaginary). Consequently, high trapping
efficiency is possible only in the regime of small $\mu$ and intermediate
$\kappa$ (see Fig. \ref{fig:ENAQT-finite}a). High ENAQT is also
only possible in this region because outside of it, loss is so dominant
that dephasing will be unable to appreciably increase trapping. This
is illustrated in Fig. \ref{fig:ENAQT-finite}b, where it can be seen
that ENAQT is large only where $\eta_{0}$ is neither very close to
0 nor very close to 1. That is, ENAQT occurs when neither trapping
nor loss is severely dominant, meaning that noise can push the balance
in favour of trapping.

\section{ENAQT on a finite chain}

\subsection{Analytical solution}

Although there is no general solution, ENAQT can be determined analytically
in every particular finite system, which is how Plenio and Huelga
proved that $\xi=0$ in the case with the origin and trap at opposite
ends of the chain \cite{Plenio:2008ff}. The solution is by Gaussian
elimination (see Appendix), meaning that $\eta$ is a rational function
of $\gamma$, $\kappa$, and $\mu$. For the three-site example in
Fig. \ref{fig:Definition-of-ENAQT}a, 
\begin{equation}
\eta=\frac{\alpha_{2}\gamma^{2}+\alpha_{1}\gamma+\alpha_{0}}{\beta_{3}\gamma^{3}+\beta_{2}\gamma^{2}+\beta_{1}\gamma+\beta_{0}},\label{eq:etadef}
\end{equation}
where $\alpha_{2}=4\kappa\mu$, $\alpha_{1}=\kappa\left(2+2\kappa\mu+8\mu^{2}\right)$,
$\alpha_{0}=\kappa\left(2\kappa\mu^{2}+\kappa+4\mu^{3}+2\mu\right)$,
$\beta_{3}=8\mu^{2}(\kappa+\mu)$, $\beta_{2}=4\mu\left(2\kappa^{2}\mu+\kappa\left(8\mu^{2}+3\right)+6\mu^{3}+4\mu\right)$,
$\beta_{1}=2\kappa^{3}\mu^{2}+2\kappa^{2}\mu\left(9\mu^{2}+5\right)+2\kappa\left(20\mu^{4}+20\mu^{2}+1\right)+6\left(4\mu^{5}+6\mu^{3}+\mu\right)$,
and $\beta_{0}=2\kappa^{3}\left(\mu^{3}+\mu\right)+\kappa^{2}\left(10\mu^{4}+13\mu^{2}+1\right)+\kappa\mu\left(16\mu^{4}+29\mu^{2}+6\right)+4\mu^{2}\left(2\mu^{4}+5\mu^{2}+2\right)$.
ENAQT is calculated by maximising this function with respect to $\gamma$.
In particular, it can be found that $\partial\eta/\partial\gamma$
can only equal zero if $-\kappa^{4}\mu+4\kappa^{3}\mu^{4}-2\kappa^{3}\mu^{2}+2\kappa^{3}+\kappa^{2}\left(20\mu^{4}+7\mu^{2}+9\right)\mu+32\kappa\mu^{6}+16\kappa\mu^{4}+7\kappa\mu^{2}-\kappa+16\mu^{7}+8\mu^{5}-4\mu^{3}-2\mu\ge0$,
meaning that there is a region in the $(\mu,\kappa)$ plane in which
ENAQT is impossible, as shown in Fig. \ref{fig:ENAQT-finite}. In
all other cases, ENAQT is strictly positive. Maximum ENAQT is $\xi_{\mathrm{max}}=7-4\sqrt{3}\approx0.0718$,
obtained as $\kappa$ and $\mu$ simultaneously tend to zero while
keeping $\mu=\kappa/2\sqrt{3}$. In that limit, $\gamma_{\mathrm{opt}}$
tends to $\sqrt{\frac{1+\sqrt{3}}{2+8\sqrt{3}}}\approx0.4151$.

\begin{figure}
\includegraphics[width=8.5cm]{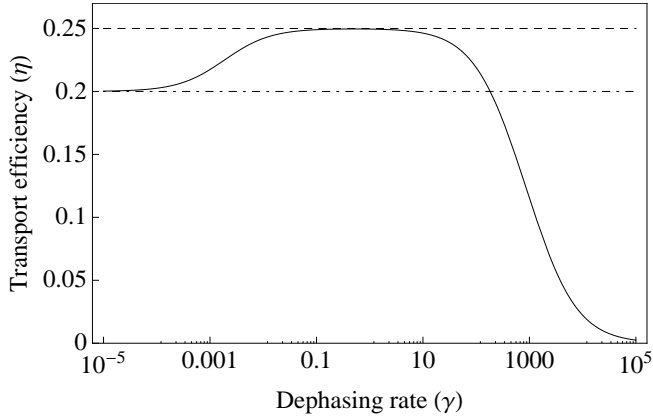}

\caption{ENAQT in the three-site system from Fig. \ref{fig:Definition-of-ENAQT}a,
in the limit of small attenuation, with $\kappa=\mu=10^{-3}$. As
$\gamma\to0$, the time evolution is merely perturbed coherent evolution,
and the efficiency approaches the predicted value of $\frac{1}{5}$
(dash-dotted line). As $\gamma\to\infty$, the Zeno effect ensures
$\eta\to0$. In the intermediate regime, $\gamma\sim1$, the dephasing
creates a fully mixed state, for which the predicted efficiency is
$\frac{1}{4}$ (dashed line), giving an ENAQT of $\frac{1}{20}$.
\label{fig:limits}}
$ $
\end{figure}

\subsection{Limit of small attenuation}

It is difficult to form a simple, intuitive picture of ENAQT in this
system that remains valid in all parameter regimes, and particularly
when time scales converge, e.g. at $\kappa\sim1$ or $\mu\sim1$.
Nevertheless, there is a simple expression for ENAQT in the limit
$\kappa,\,\mu\ll1$. In that case, both attenuation mechanisms are
weak and can be treated as perturbations on the dephased quantum dynamics.
In particular, because attenuation is slow compared to the quantum
dynamics, we assume that we can only consider the average site populations
in calculating loss and trapping. In the case with appreciable dephasing,
the state of the system will quickly reach a completely mixed state,
meaning that each site will host $1/N$ of the remaining population.
In particular, the rate at which the particle will be trapped at the
trap site will equal $\kappa/N$. Similarly, all population will be
lost at a rate $\mu$, giving the efficiency 
\begin{equation}
\eta_{\mathrm{dephased}}\approx\frac{\kappa/N}{\kappa/N+\mu}=\frac{1}{1+N\mu/\kappa}.
\end{equation}
In the coherent case, the system eigenstates are $u_{j}^{(k)}=\sqrt{\frac{2}{N+1}}\sin\frac{\pi jk}{N+1}$
with eigenvalues $\lambda_{k}=2\cos\frac{\pi k}{N+1}$. Therefore,
the amplitude of site $l$ given an initial site $m$ is 
\begin{equation}
U_{lm}(t)=\frac{2}{N+1}\sum_{j=1}^{N}\sin\frac{\pi jl}{N+1}\sin\frac{\pi jm}{N+1}e^{-it\cdot2\cos\frac{\pi j}{N+1}},
\end{equation}
which can be used to show that the average population $\overline{P}_{lm}=\lim_{T\to\infty}\frac{1}{T}\int_{0}^{T}|U_{lm}(t)|^{2}dt$
equals
\begin{equation}
\overline{P}_{lm}=\frac{1}{N+1}\left(1+\frac{1}{2}\delta_{lm}+\frac{1}{2}\delta_{l,N+1-m}\right),
\end{equation}
where $\delta_{lm}$ is the Kronecker delta function. From there we
have $\eta_{\mathrm{coherent}}\approx\left(1+\overline{P}_{lm}^{-1}\mu/\kappa\right)^{-1}$.
For ENAQT, we do not consider the case $l=m$, meaning that there
are two situations, depending on whether the trap is opposite the
initial site. Because partial recurrences can refocus excitation from
the initial site to the opposite site, if the opposite site is the
target, $l=N+1-m$, the average target population exceeds $\frac{1}{N}$
and ENAQT is impossible. This explains, at least in the limit of small
$\kappa$ and $\mu$, Plenio and Huelga's observation of the absence
of ENAQT in end-to-end transfer. In transport between sites that are
not opposite each other, 
\begin{equation}
\xi\approx\frac{1}{1+N\mu/\kappa}-\frac{1}{1+(N+1)\mu/\kappa}>0.\label{eq:analyticenaqt}
\end{equation}
Notably, $\xi$ depends only on the ratio $\mu/\kappa$. The validity
of these approximations is demonstrated in Fig. \ref{fig:limits}.
The expression is more accurate for large $\mu/\kappa$ because loss,
by lowering all the amplitudes simultaneously, perturbs the time evolution
less than trapping, which affects only one site.

\subsection{Other patterns}

Several patterns emerge in longer chains and when the locations of
the trapping site and the initial site are varied. Table \ref{tab:Max-ENAQT}
shows the maximum possible ENAQT in chains up to $N=8$ with all possible
combinations of initial and trap sites. Each entry is calculated by
analytically solving the equations of motion and maximising $\xi$
as a function of $\kappa$ and $\mu$, as discussed above for the
$N=3$ chain. As we proved above, we can see that ENAQT is possible
in all configurations except when the initial site is located opposite
the trap. Furthermore, maximum ENAQT increases with increasing $N$.
This is the opposite of the trend predicted by Eq. \ref{eq:analyticenaqt},
and occurs because the high values observed in the table generally
occur in the regime of small $\mu/\kappa$, where the estimate of
$\eta_{\mathrm{coherent}}$ fails.

\begin{table}[t]
{\small
\begin{tabular}{ccllllllll}
\toprule &  & \multicolumn{8}{c}{Initial site}\tabularnewline
\cmidrule{3-10}$N$ & $\tau$ & 1 & 2 & 3 & 4 & 5 & 6 & 7 & 8\tabularnewline
\midrule3 & 1 & X & 0.072 & 0 &  &  &  &  & \tabularnewline
3 & 2 & $1/2$ & X & $1/2$ &  &  &  &  & \tabularnewline
\midrule4 & 1 & X & 0.083 & 0.083 & 0 &  &  &  & \tabularnewline
4 & 2 & 0.083 & X & 0 & 0.083 &  &  &  & \tabularnewline
\midrule5 & 1 & X & 0.082 & 0.107 & 0.082 & 0 &  &  & \tabularnewline
5 & 2 & $1/3$ & X & $1/3$ & 0 & $1/3$ &  &  & \tabularnewline
5 & 3 & $1/2$ & $1/2$ & X & $1/2$ & $1/2$ &  &  & \tabularnewline
\midrule6 & 1 & X & 0.080 & 0.114 & 0.114 & 0.080 & 0 &  & \tabularnewline
6 & 2 & 0.114 & X & 0.080 & 0.080 & 0 & 0.114 &  & \tabularnewline
6 & 3 & 0.080 & 0.114 & X & 0 & 0.114 & 0.080 &  & \tabularnewline
\midrule7 & 1 & X & 0.077 & 0.115 & 0.125 & 0.115 & 0.077 & 0 & \tabularnewline
7 & 2 & $1/4$ & X & $1/4$ & 0.033 & $1/4$ & 0 & $1/4$ & \tabularnewline
7 & 3 & 0.115 & 0.077 & X & 0.125 & 0 & 0.077 & 0.115 & \tabularnewline
7 & 4 & $1/2$ & $1/2$ & $1/2$ & X & $1/2$ & $1/2$ & $1/2$ & \tabularnewline
\midrule8 & 1 & X & 0.074 & 0.114 & 0.128 & 0.128 & 0.114 & 0.074 & 0\tabularnewline
8 & 2 & 0.128 & X & 0.114 & 0.074 & 0.074 & 0.114 & 0 & 0.128\tabularnewline
8 & 3 & $1/3$ & $1/3$ & X & $1/3$ & $1/3$ & 0 & $1/3$ & $1/3$\tabularnewline
8 & 4 & 0.074 & 0.128 & 0.114 & X & 0 & 0.114 & 0.128 & 0.074\\ \bottomrule\tabularnewline
\end{tabular}
}

\caption{Analytically calculated maximum ENAQT in a chain with $N$ sites,
depending on the initial and trap ($\tau$) sites. ENAQT is not defined
if the initial and trap sites coincide (X). All the values with the
same $N$ that are equal to three decimal places are exactly equal.
$0.072=7-4\sqrt{3}$ and the rest are roots of more complicated, but
known, polynomials.\label{tab:Max-ENAQT}}
\end{table}

It can also be seen that, regardless of the trap site, $\xi_{\mathrm{max}}$
is equal for situations with initial sites $m$ and $N+1-m$. This
is the case even though $\xi(\mu,\kappa)$ is not equal in the two
situations in general. In the limit of infinitesimally small $\kappa$,
$\mu$, and $\gamma$, where the equality obtains, coherent time evolution
proceeds before any appreciable dephasing or loss takes places. Therefore,
coherent recurrences can occur, and although the recurrences are imperfect
because the eigenvalues $\lambda_{k}$ are incommensurable, they become
arbitrarily close to perfect after a sufficiently long time. In particular,
a particle initialised at $m$ will refocus (arbitrarily close to
perfectly) at $N+1-m$ after sufficient time. On the longer time scale
of loss, the two initial conditions become indistinguishable.

\subsection{High ENAQT in symmetric situations\label{sub:sym}}

As shown in Table \ref{tab:Max-ENAQT}, a high ENAQT of $\xi_{\mathrm{max}}=\frac{1}{2}$
occurs if the trap is in the middle of a chain with an odd number
of sites, regardless of the initial site. In that case, the full Hamiltonian
$H$ commutes with the inversion operator $P$, defined as $P\left|j\right\rangle =\left|N+1-j\right\rangle $.
The initial site $\left|\tau\right\rangle $ can be written as an
equal superposition of symmetric and antisymmetric states, $\left|\tau\right\rangle =\frac{1}{\sqrt{2}}\left(\left|S\right\rangle +\left|A\right\rangle \right)$,
where$\left|S\right\rangle =\frac{1}{\sqrt{2}}\left(\left|\tau\right\rangle +\left|N+1-\tau\right\rangle \right)$
and $\left|A\right\rangle =\frac{1}{\sqrt{2}}\left(\left|\tau\right\rangle -\left|N+1-\tau\right\rangle \right)$.
Because $\left|A\right\rangle $ is odd, $P\left|A\right\rangle =-\left|A\right\rangle $,
and $H$ commutes with $P$, $\left|A\right\rangle $ remains odd
under time evolution. Since the trap site is in the middle, it is
even, meaning that the $\left|A\right\rangle $ component of $\left|\tau\right\rangle $
never gets mapped to the trap site and can therefore not be trapped.
By contrast, the $\left|S\right\rangle $ component does get trapped.
Therefore, the efficiency at zero dephasing is $\eta_{0}=\frac{1}{2}$.
When dephasing is non-zero, the phase coherence in $\left|A\right\rangle $
is lost, meaning that now the particle can be completely trapped.
In particular, if $\kappa\gg\mu$, a negligible amount will be lost,
meaning that $\eta_{\mathrm{max}}=1$, giving $\xi=\eta_{\mathrm{max}}-\eta_{0}=\frac{1}{2}$.

\section{ENAQT on a circle}

If the transport takes place on a circle instead of a finite chain,
the behaviour is qualitatively the same. We consider here the same
situation as above, except that in Eq. \ref{eq:H}, there is an additional
term coupling sites $\left|1\right\rangle $ and $\left|N\right\rangle $.

In the regime of weak attenuation, $\kappa,\mu\ll1$, as in the chain,
$\eta_{\mathrm{dephased}}\approx\left(1+N\mu/\kappa\right)^{-1}$.
In the coherent case, the eigenstate amplitudes are $u_{j}^{(k)}=\frac{1}{\sqrt{N}}e^{2\pi ijk/N}$
with eigenvalues $\lambda_{k}=2\cos\frac{2\pi k}{N}$, from where
the average population is
\begin{equation}
\overline{P}_{lm}=\begin{cases}
\frac{1}{N^2} \left(N \left( 1+\delta_{lm} \right) -1\right) , & N\text{ odd}\\
\frac{1}{N^2} \left(N \left(1+\delta_{lm}+\delta_{l,m+N/2}\right)-2\right), & N\text{ even.}
\end{cases}
\end{equation}
Because $\overline{P}_{lm}>\frac{1}{N}$ if $l=m+\frac{N}{2}$, there
is no ENAQT if the initial site lies on the opposite side of the circle
to the trap. In other cases, 
\begin{equation}
\xi\approx\frac{1}{1+N\mu/\kappa}-\frac{1}{1+N^{2}\mu/(N-1)\kappa}>0.\label{eq:analyticenaqt-circle}
\end{equation}
As in the chain, this expression is more accurate for large $\mu/\kappa$.

Regardless of the number of sites, maximum possible ENAQT on the circle
is
\begin{equation}
\xi_{\mathrm{max}}=\begin{cases}
0, & \text{initial and trap sites opposite}\\
\frac{1}{2}, & \text{otherwise.}
\end{cases}
\end{equation}
The situation is much simpler than in the chain, where Table \ref{tab:Max-ENAQT}
is needed. The high value of $\xi_{\mathrm{max}}$ occurs because
the circle always has the inversion symmetry required in Sec. \ref{sub:sym}.

\begin{figure*}
\includegraphics[width=10cm]{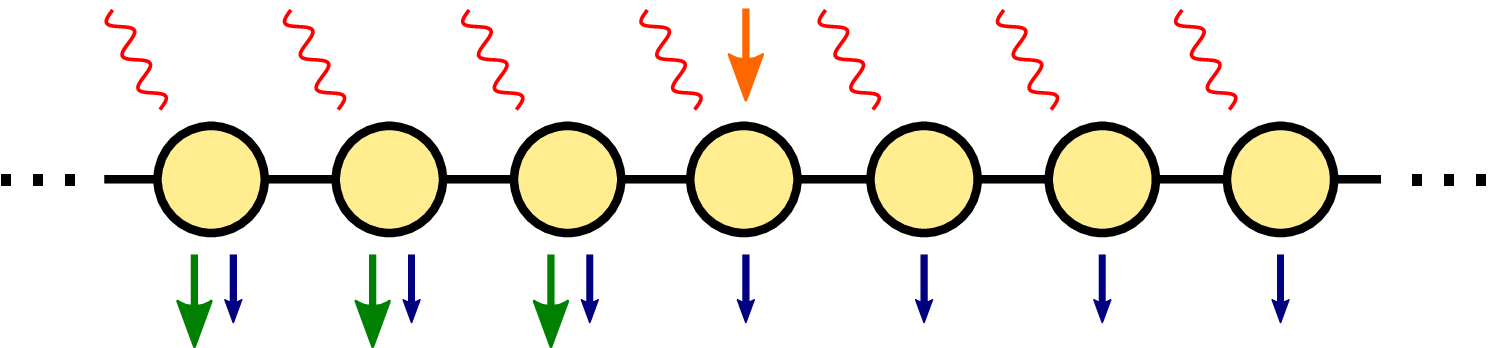}

\bigskip{}

{\footnotesize Efficiency without dephasing ($\eta_{0}$)\hfill ENAQT ($\xi$)\hfill Optimal
dephasing ($\gamma_{\mathrm{opt}}$)}

\includegraphics[width=1\textwidth]{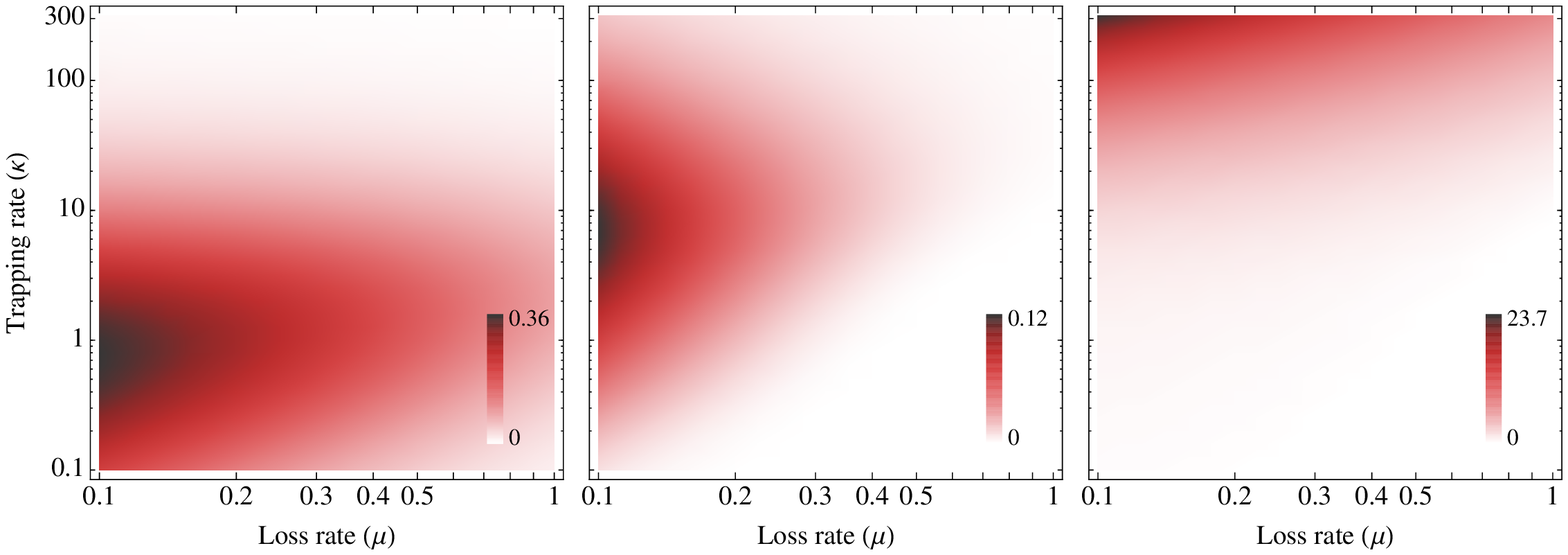}

\caption{The infinite system (\textbf{top}) and its characterisation (\textbf{bottom}).
As in the finite chain (Fig. \ref{fig:ENAQT-finite}), high $\eta_{0}$
occurs for small loss and intermediate trapping, ENAQT is highest
at intermediate $\eta_{0}$, and the dephasing required for ENAQT
grows rapidly with increasing $\kappa$. \label{fig:ENAQT-in-infinite}}
\end{figure*}

\section{ENAQT on an infinite chain}

ENAQT also occurs---albeit by a different mechanism---in infinite
ordered systems. The dephased dynamics of a particle on an infinite
chain is well understood in the absence of trapping and loss \cite{Schwarzer:1972br,Kendon:2007fw,Hoyer:2010fl}.
In the fully coherent case, $\gamma=0$, the dynamics is ballistic,
while increasing noise converts it to diffusion.

To observe ENAQT, we introduce loss $\mu$ everywhere and trapping
$\kappa$ on all sites to the left of the initial site (see Fig. \ref{fig:ENAQT-in-infinite}a).
In the coherent case, there is a sizeable probability that the particle,
owing to its ballistic motion, will move far to the right, eventually
being lost. In the opposite extreme, $\gamma=\infty$, the particle
is completely localised at the initial site by the Zeno effect, and
therefore eventually lost. In the intermediate region, ENAQT is possible
because decoherence slows down the spreading sufficiently to prevent
the rightward-moving component from escaping, but not strongly enough
to prevent the particle from diffusing into the trap region.

Fig. \ref{fig:ENAQT-in-infinite}b shows numerically computed ENAQT
on the infinite chain.  The infinite line was represented by sufficiently
many sites $N$ to avoid the particle reaching the edges. Because
the simulation time and $N$ both scale as $\mu^{-1}$, the computation
becomes expensive for small $\mu$, and we have imposed a lower cutoff
of $\mu=0.1$. The largest ENAQT found is $\xi(\mu=0.1,\,\kappa=6.3)=0.1233$.
Similar results are obtained with initial sites further from the trapping
region. In those cases, ENAQT is smaller because the particle has
to travel farther to the trapping region, but it remains finite for
small $\mu$ and intermediate $\kappa$. ENAQT tend to zero as $\mu\to0$,
as $\mu\to\infty$, and as $\kappa\to\infty$ for the same reasons
as in the finite chain. A question we are not able to answer with
numerical simulations is whether ENAQT is ever exactly zero or merely
approaches zero asymptotically in the appropriate limits.

\section{Conclusions}

We have shown at least two different mechanisms for ENAQT in an ordered
system. In the finite lattice with small $\kappa$ and $\mu$, it
is caused by the fact that a dephasing-induced mixed state is more
likely to be found at the trap site than a coherently propagated initial
state, except if the initial and trap sites are opposite each other.
In the infinite lattice, it occurs when dephasing slows down otherwise
ballistic transport and prevents a portion of the particle from escaping
far from the trap region. We leave open the questions of whether the
various mechanisms of ENAQT (including the ones in disordered systems)
can be understood in a unified picture and how they are influenced
by different kinds of noise, of which pure dephasing is only one limit.

\begin{ack}
We thank Gerard Milburn, Patrick Rebentrost, and Andrew White for
valuable discussions, and Martin Plenio for assistance with the material
in the Appendix. We acknowledge support from a UQ Postdoctoral Research
Fellowship (IK), DARPA (N66001-10-1-4063), and the Camille and Henry
Dreyfus Foundation (AAG).
\end{ack}

\section*{Appendix: Analytical calculation of the efficiency}

\setcounter{equation}{0}
\renewcommand{\theequation}{A\arabic{equation}}The initial state of the system $\rho(0)$ can be understood as a
vector in a Liouville space of dimension $N^{2}$. In order to calculate
the efficiency, we augment it to $\tilde{\rho}(0)$ with $N^{2}+1$
entries. The first $N^{2}$ entries we call the state sector, and
they equal $\rho(0)$, while the final entry, the accumulator, we
initialised to $\tilde{\rho}_{N^{2}+1}(0)=0$. The Liouvillian $\mathcal{L}$
is likewise modified to $\tilde{\mathcal{L}}$, an $\left(N^{2}+1\right)\times\left(N^{2}+1\right)$
matrix, where the top-left $N^{2}\times N^{2}$ elements equal $\mathcal{L}$,
and the remainder are set to 0, except for the entry $\tilde{\mathcal{L}}_{N^{2}+1,\tau'}=2\kappa$,
where $\tau'=1+(\tau-1)(N+1)$ is the coordinate of the population
of the trap site $\tau$. That is, $\tilde{\mathcal{L}}$ couples
the population of the trap site to the accumulator (but not vice-versa)
with strength $2\kappa$. Because $\tilde{\mathcal{L}}$ does not
couple from the accumulator to the state sector of $\tilde{\rho}(t)$,
the time evolution of the state sector under $\tilde{\mathcal{L}}$
is equal to the time evolution of $\rho(t)$ under $\mathcal{L}$.
During the evolution, the accumulator increases precisely at the rate
$2\kappa\rho_{\tau'}(t)$, meaning that $\tilde{\rho}_{N^{2}+1}(\infty)=\eta$,
while the remaining elements of $\tilde{\rho}(\infty)$ are all reduced
to 0. In principle, one could calculate $\eta$ by calculating $\tilde{\rho}(\infty)=\lim_{t\to\infty}e^{\tilde{\mathcal{L}}t}\tilde{\rho}(0)$,
but this appears to us to be too difficult analytically. 

Instead of solving the initial-value problem, we solve a related steady-state
equation. We begin by modifying $\tilde{\mathcal{L}}$ to $\tilde{\mathcal{L}}^{\varepsilon}$,
which is the same except for $\tilde{\mathcal{L}}_{N^{2}+1,N^{2}+1}^{\varepsilon}=\varepsilon$.
Then we solve, analytically by Gaussian elimination or otherwise,
the linear system of equations
\begin{equation}
\tilde{\mathcal{L}}^{\varepsilon}\tilde{\sigma}=\varepsilon\tilde{\rho}(0).\label{eq:ss}
\end{equation}
Here, $\tilde{\sigma}$ is the steady state in the situation where
$\rho(0)$ is being injected into the system at rate $\varepsilon$.
In particular, since total probability being injected into the system
is $\varepsilon$, the fraction $\varepsilon\eta$ must go to the
accumulator. Since the accumulator accumulates at a rate $2\kappa\tilde{\sigma}_{\tau'}$,
we must have $\tilde{\sigma}_{\tau'}=\varepsilon\eta/2\kappa$. Now,
the accumulator component of Eq. \ref{eq:ss} is $2\kappa\tilde{\sigma}_{\tau'}+\varepsilon\tilde{\sigma}_{N^{2}+1}=0$,
from where we can conclude that $\tilde{\sigma}_{N^{2}+1}=\eta$,
meaning that the efficiency can be read out of the solution $\tilde{\sigma}$.
The result is independent of $\varepsilon$, meaning that the procedure
remains valid in the limit of infinitesimally small $\varepsilon$,
where Eq. \ref{eq:ss} reduces to a true steady-state equation, $\dot{\tilde{\sigma}}=0$

\section*{References}
\bibliographystyle{unsrt}

\end{document}